\documentclass[journal]{IEEEtran}
\usepackage{cite}
\usepackage{amsfonts}
\usepackage{amsmath}
\usepackage{balance}
\ifCLASSINFOpdf
\else
   \usepackage[dvips]{graphicx}
\fi
\usepackage{url}
\usepackage{graphicx}
\usepackage[multiple]{footmisc}

\usepackage{url}          
\usepackage{doi}
\usepackage{hyperref}

\usepackage{booktabs}
\usepackage{graphicx}
\usepackage{booktabs}
\usepackage{amsmath}
\usepackage{array}
\usepackage{makecell}
\usepackage{multirow} 
\usepackage{tabularx}
\usepackage{adjustbox}
\usepackage{amsmath,amssymb}

\usepackage{threeparttable}

\begin{document}

\title{Noise-Robust Sound Event Detection and Counting via Language-Queried Sound Separation}

\author{
Yuanjian Chen*\thanks{*Corresponding author (email:2010400002@stu.hrbust.edu.cn)}, 
Yang Xiao, \IEEEmembership{Graduate Student Member, IEEE},
Han Yin, 
Yadong Guan, \IEEEmembership{Member, IEEE},
and Xubo Liu, \IEEEmembership{Member, IEEE}
}

\markboth{Journal of \LaTeX\ Class Files, Vol. 14, No. 8, August 2015}
{Shell \MakeLowercase{\textit{et al.}}: Bare Demo of IEEEtran.cls for IEEE Journals}
\maketitle

\begin{abstract}
Most sound event detection (SED) systems perform well on clean datasets but degraded significantly in noisy environments. Language-queried audio source separation (LASS) models show promise for robust SED by separating target events, existing methods require elaborate multi-stage training and lack explicit guidance for target events. To address these challenges, we introduce event appearance detection (EAD), a counting‐based manner that counts event occurrences at both the clip and frame levels. Based on EAD, we introduce a co-training–based multi-task learning framework for EAD and SED to enhance SED’s performance in noisy environments. First, SED struggles to learn the same pattern as EAD. Then, a task-based constraint is designed to improve prediction consistency between SED and EAD. This framework provides more reliable clip-level predictions for LASS models and strengthen timestamps detection capability. Experiments on DESED and WildDESED datasets demonstrate better performance compared to existing methods, with advantages becoming more pronounced at higher noise levels. The code is available at: https://github.com/visionchan/EADSED.
\end{abstract}


\IEEEpeerreviewmaketitle

\section{Introduction}
\IEEEPARstart{R}{ecent} progress in computational auditory scene analysis (CASA) \cite{wang2006computational} has been driven by artificial intelligence. One important task of CASA is sound event detection (SED) \cite{cakir2017convolutional,gao2024local,cai2025prototype}, which aims to find specific sound events and mark their starting and ending times. SED has been growing rapidly and is now applied in many real-world areas, such as urban safety monitoring \cite{tan2022comprehensive}, environmental research \cite{bahmei2022cnn}, and medical applications \cite{tran2022multi,alqudaihi2021cough,qiu2024heart}.
 
Many important studies \cite{cakir2017convolutional,kong2020sound,nam2022frequency,shao2024fine,xiao2025semi} have focused on learning distinctive sound representations for known sound categories in clean environments. However, real-world environments are often more complex. In these settings, background noise often appears \cite{xiao2024wilddesed}. These noises are not marked as target events, and they often mix with the sounds of interest. Because of this, systems trained exclusively on clean data usually perform poorly in noisy environments.

Studies have shown that noisy environments can significantly reduce the performance of SED systems that are trained on clean audio \cite{foggia2015reliable,neri2022sound}. This emphasizes the need for models that are robust to noise. Current noise-robust SED methods generally fall into two categories: fine-tuning with noisy data \cite{foggia2015reliable,neri2022sound,mcloughlin2015robust,ozer2018noise,bhosale2024diffsed} to improve generalization, and multi-task learning \cite{kong2019sound,liang22_interspeech,ucil} that incorporates auxiliary tasks such as sound separation \cite{turpault2021sound,della2024resource}, often linked to sound source counting \cite{he2024soundcount}. Sound event counting has also been used in other domains, such as repetitive action counting \cite{zhang2021repetitive,lee2024multimodal}, to address incomplete information in a single modality. However, most counting-based methods overlook challenging scenarios with heavy polyphony and overlapping events. Some studies \cite{he2024soundcount,guan2024sound} estimate event counts at a single granularity either frame-level or clip-level and often rely on closed-set assumptions of known event categories. This limits their ability to handle numerous unknown noise events in real-world audio. Meanwhile, language-queried audio source separation (LASS) \cite{liu2024separate} has shown promise in improving SED performance in noisy conditions \cite{11011161}, but it depends heavily on accurate text queries. In high-noise or highly concurrent scenarios, SED outputs may be incomplete or incorrect, leading to suboptimal separation. This reveals two key challenges: (1) how to maintain SED robustness as noise intensity and event concurrency increase, and (2) how to provide more accurate queries for LASS during inference. Efficiently integrating the strengths of counting tasks with LASS is therefore critical.

To address these challenges, we propose a new auxiliary task, event appearance detection (EAD), which bridges counting and LASS within a unified SED framework. EAD operates at two levels: global-EAD predicts whether a clip contains one, two, or more than two events, and local-EAD identifies whether a frame contains no event, a single event, or multiple events. Unlike conventional methods that rely solely on text-based queries, EAD uses event counts as an internal supervisory signal to guide SED learning. This category-independent design makes it more generalizable and less sensitive to unseen noise types. By enforcing consistency constraints between EAD and SED outputs, our framework learns shared noise-robust features, enabling both more accurate timestamp predictions and improved text queries for LASS.

As we mentioned above, to adapt to the noisy environment, our framework leverages multi-task learning to train SED and EAD, enabling them to support each other. During testing, only the SED branch is active, and clip-level outputs are used as text queries for the LASS model. The LASS-separated sounds are then refined by the SED model for final detection. Compared to previous methods like \cite{11011161}, our system enhances noise robustness, removes the need for multi-stage training, and requires no extra manual labels, as EAD annotations are automatically derived from existing audio tagging and SED labels. Experiments on the DESED and WildDESED datasets confirm that our framework outperforms single-task baselines and presents a practical and scalable solution for real-world SED in noisy conditions. The code is publicly available. 

\begin{figure*}[t]
\centering
\includegraphics[width=0.8\textwidth]{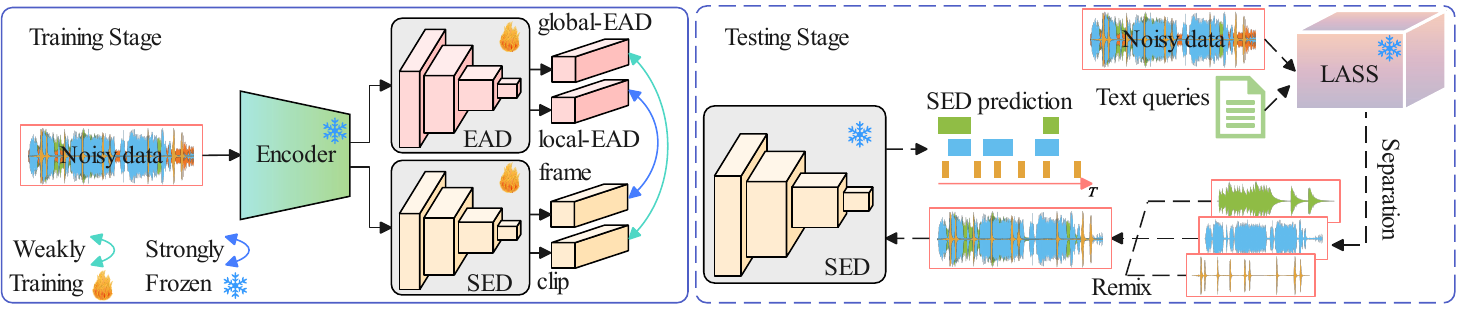}
\vspace{-7pt}
\caption{Overview of the proposed cooperative dual-branch framework for SED in noisy environments. During the training stage, audio is processed by a shared encoder (BEATs), followed by two parallel branches: the SED branch (for frame-level and clip-level predictions) and the EAD branch (for global and local EAD). In the testing stage, only the SED branch is used. Clip-level predictions are converted into text queries for the LASS model, which separates sound sources. The separated signals are then remixed and passed back to the SED model for final prediction.}
\vspace{-12pt}
\label{fig:fig1}
\end{figure*}

\section{Proposed cooperative framework}

This section presents our cooperative dual-branch framework. The framework includes two branches: the SED branch and the EAD branch, along with a collaborative training and testing process, as illustrated in Fig.\ref{fig:fig1}. This design ensures accurate frame-level timestamps predictions and produces reliable clip-level text queries for the LASS model during testing.

\subsection{Sound Event Detection Branch}

We use a standard convolutional neural network (CRNN) architecture \cite{cakir2017convolutional} as the base feature extractor for SED in noisy environments, without any special modifications. The CRNN outputs frame-level predictions $Z_s \in \mathbb{R}^{T \times C}$, where $T$ is the number of time frames and  $C$ is the number of sound event classes. Then, an attention pooling mechanism aggregates these frame-level outputs into clip-level predictions $Z_w \in \mathbb{R}^{C}$. Both outputs $Z_s$ and $Z_w$ are supervised using binary cross entropy loss functions. The frame-level loss is calculated based on the strong labels $Y_s \in \{0,1\}^{T \times C}$, and the clip-level loss uses the weak labels $Y_w \in \{0,1\}^{C}$. The overall SED loss is thus defined as:

\begingroup
\vspace{-4mm}
\begin{equation}
\mathcal{L}_{\text{SED}} = \mathcal{L}_s(Z_s, Y_s) + \mathcal{L}_w(Z_w, Y_w)
\label{eq:L_sed}
\end{equation}
\vspace{-6mm}
\endgroup

This setup enables the model to learn both detailed temporal information and global presence of events.

\subsection{Event Appearance Detection Branch}

\begin{figure}[t!]
\centering
\includegraphics[width=0.68\columnwidth]{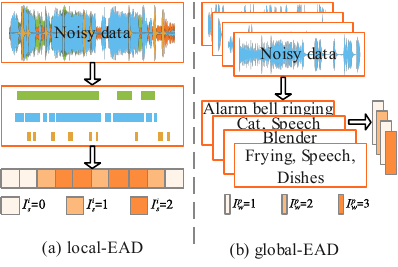}
\vspace{-15pt}
\caption{Overview of local-EAD and global-EAD methods. Local-EAD uses strong labels to classify each frame as containing no, one, or multiple events. Global-EAD uses weak labels to label clips based on the number of unique events: one, two, or more than two.}
\label{fig:fig2}
\vspace{-13pt}
\end{figure}

In real domestic environments, sound events often overlap with background noise due to the complex and dynamic nature of such settings. This overlap makes it difficult for a standard SED system to accurately extract the distinct characteristics of target events, such as their frequency patterns \cite{guan2024sound}. Moreover, our goal is to improve timestamps localization of sound events using a single-stage framework that shares the same model for all tasks, which differs from previous multi-stage strategies like in \cite{11011161}. To achieve this, we introduce event appearance detection (EAD), designed to enhance SED robustness in noisy environments. The goal of EAD is to model the activity level of target sound events, rather than recognizing specific event types. It captures the number of active events at both the frame and clip levels. During training, the EAD branch shares the same architecture as the SED branch. However, during testing, the EAD branch remains inactive, which means it does not introduce any extra computational cost during testing stage.

As illustrated in Fig.\ref{fig:fig2}, at the frame-level, we adopt the idea of \cite{guan2024sound} to design local-EAD, which classifies each frame into one of three categories: no event, a single event, or multiple overlapping events. For strongly labeled data, we define the local-EAD label as $\mathcal{I}_s \in \{0,1,2\}^{T}$ for a given clip. Since local-EAD aligns temporally with SED, we compute $\mathcal{I}_s$ by summing the strong labels across all classes and applying a cap of 2 using $\min\{2,\bullet \}$, as shown in Equation(\ref{eq:Iis}). This aggregation not only reduces the impact of noisy or unlabeled sound events on the local-EAD method but also enables it to co-train with the SED task, improving timestamps precision.

\begingroup
\vspace{-3mm}
\begin{equation}
\mathcal{I}_s^i = \min\left\{2, \sum_{c} Y_s^{i,c}\right\},\quad c \in \{1,2,\ldots,C\}
\label{eq:Iis}
\end{equation}
\vspace{-3mm}
\endgroup

For training, we convert $\mathcal{I}_s$ to a one-hot encoded format, $\bar{\mathcal{I}_s}\in \{0,1\}^{T\times 3}$, and define the local-EAD model's output as $\Pi_s \in \mathbb{R}^{T\times 3}$. The loss for local-EAD is computed using cross entropy loss: \(\mathcal{L}_{\text{local}} = CE(\Pi_s, \bar{\mathcal{I}_s})\).

At the clip-level, we define global-EAD, which classifies an entire audio clip into three categories: containing one event, two events, or more than two events. For weakly labeled data, we represent the global-EAD label as $\mathcal{I}_w \in \{1,2,3\}$. Similar to the local version, we apply a one-hot encoding $\bar{\mathcal{I}_w} \in\{0,1\}^{1 \times 3}$, and the prediction output is $\Pi_w \in \mathbb{R}^{1 \times 3}$. The global loss is: \(\mathcal{L}_{\text{global}} = CE(\Pi_w, \bar{\mathcal{I}_w})\), where \(\mathcal{I}_w^p\) is:

\begin{equation}
\mathcal{I}_w^p = \min\left\{3, \sum_{c} Y_w^{p,c}\right\},\quad c \in \{1,2,\ldots,C\}
\label{eq:Ipw}
\end{equation}

Together, the total EAD loss is defined as the sum of \(\mathcal{L}_{\text{local}}\) and \(\mathcal{L}_{\text{global}}\). Because EAD focuses only on the number of active events, its structure is simpler and lighter than SED. When optimized together, the cooperative learning of EAD and SED provides mutual benefits, strengthening both frame-level timestamps detection and clip-level presence estimation, especially under noisy conditions.

\subsection{Cooperative Training and Noise-robust Testing}

During training, we adopt the Mean Teacher (MT) method \cite{tarvainen2017mean} for semi-supervised learning, following previous work such as \cite{11011161}. The total supervised loss in MT, denoted as $\mathcal{L}_{\text{SUP}}$, includes both the SED loss and the EAD loss. A hyperparameter $\rho_{\text{SUP}}$ controls the EAD component:

\begingroup
\vspace{-5mm}
\begin{equation}
\mathcal{L}_{\text{SUP}} = \mathcal{L}_{\text{SED}} + \rho_{\text{SUP}} \mathcal{L}_{\text{EAD}}
\label{eq:L_SUP}
\end{equation}
\vspace{-7mm}
\endgroup

We also introduce inter-task consistency constraints. The key idea is that the number of events predicted by the SED branch should align with the event counts estimated by the EAD branch. To measure this alignment, we transform the SED outputs into event count forms at both frame and clip levels,  denoted as $\tilde{\mathcal{I}_s}$ and $\tilde{\mathcal{I}_w}$ respectively to match the probabilistic outputs of the EAD model.
For frame-level consistency, we compute a difference score \(\Theta_{s}^{i}\) for the \(i\)-th frame, based on the absolute difference between the predicted count from SED and the expected count from local-EAD:

\begingroup
\vspace{-4mm}
\begin{equation}
\Theta_{s}^{i}=\left | \tilde{\mathcal{I}_{s}^{i}}-E_{\bar{\mathcal{I}_{s}^{i}}}\left \{\mathcal{I}_s \right \} \right |    
\label{eq:theta_s}
\end{equation}
\vspace{-6mm}
\endgroup

\noindent where \(E\) presents the expectation of prediction. Similarly, at the clip-level, the consistency score \(\Theta_{w}^{p}\) for the \(p\)-th clip is:

\begingroup
\vspace{-3mm}
\begin{equation}
\Theta_{w}^{p}=\left | \tilde{\mathcal{I}_{w}^{p}}-E_{\bar{\mathcal{I}_{w}^{p}}}\left \{\mathcal{I}_w \right \} \right |    
\label{eq:theta_w}
\end{equation}
\vspace{-5mm}
\endgroup

Smaller values of $\Theta_{s}^{i}$ and $\Theta_{w}^{p}$ indicate stronger agreement between the SED and EAD predictions. Based on this, we define the inter-task consistency loss \(\mathcal{L}_{\text{ACC}}\):

\begingroup
\vspace{-3mm}
\begin{equation}
\mathcal{L}_{\text{ACC}}=\frac{1}{2T}\sum_{i=1}^{T}(\Theta_{s}^{i})^{2}+\frac{1}{2P}\sum_{p=1}^{P}(\Theta_{w}^{p})^{2}
\label{eq:L_ACC}    
\end{equation}
\vspace{-4mm}
\endgroup

\noindent where $P$ denotes the number of weakly labeled clips in a batch. To incorporate consistency between the teacher and student models in MT, we define the consistency loss $\mathcal{L}_{\text{CON}}$, which includes both the SED and EAD branches:

\begingroup
\vspace{-4mm}
\begin{equation}
\mathcal{L}_{\text{CON}}=\mathcal{L}_{\text{SED-CON}}+\mathcal{L}_{\text{EAD-CON}}+\rho_{\text{CON}}\mathcal{L}_{\text{ACC}}
\label{eq:L_CON}
\end{equation}
\vspace{-6mm}
\endgroup

Here, \(\mathcal{L}_{\text{SED-CON}}\) and \(\mathcal{L}_{\text{EAD-CON}}\) follow standard consistency regularization formulations as in~\cite{gao2024local}, and $\rho_{\text{CON}}$ controls the importance of the inter-task constraint. The final total loss is computed as: \(\mathcal{L}_{\text{TOTAL}}=\mathcal{L}_{\text{SUP}}+\omega \mathcal{L}_{\text{CON}}\), where $\omega$ is a ramp-up function \cite{tarvainen2017mean} that gradually increases the weight of the consistency term during training.

As shown in Fig.~\ref{fig:fig1}, the testing pipeline follows a similar structure to \cite{11011161}. However, a key advantage of our method is its efficiency. While previous approaches require multiple models for training and testing, our framework uses the same model in both stages, which reduces the number of parameters by half and simplifies deployment.

\begin{table}[t!]
\centering
\begin{threeparttable}

\caption{Ablation studies of different training objectives}

\label{tab:ablation}
\renewcommand{\arraystretch}{1.05}
\scriptsize
\begin{tabular}{@{}p{2.8cm}@{\hspace{3pt}}c@{\hspace{6pt}}*{5}{c}@{}}
\toprule
\multirow{2}{2.0cm}{Loss function} & \multirow{2}{*}{Metric} & \multicolumn{4}{c}{WildDESED} & \multirow{2}{*}{AVG.\tnote{b}} \\
\cmidrule(lr){3-6}
& & -5dB & 0dB & +5dB & +10dB & \\
\midrule
\multirow{2}{2.8cm}{$\mathcal{L}_{\text{BASELINE}}$\tnote{a}} 
& PSDS1↑ & 0.088 & 0.130 & 0.213 & 0.306 & 0.184 \\
& PSDS2↑ & 0.296 & 0.394 & 0.441 & 0.572 & 0.426 \\
\midrule
\multirow{2}{2.8cm}{$\mathcal{L}_{\text{BASELINE}} + \mathcal{L}_{\text{local}}$} 
& PSDS1↑ & 0.148 & 0.193 & 0.231 & 0.317 & 0.222 \\
& PSDS2↑ & 0.394 & 0.479 & 0.531 & 0.606 & 0.503 \\
\midrule
\multirow{2}{2.8cm}{$\mathcal{L}_{\text{BASELINE}} + \mathcal{L}_{\text{EAD}}$} 
& PSDS1↑ & 0.145 & 0.200 & \textbf{0.272} & \textbf{0.338} & 0.239 \\
& PSDS2↑ & 0.380 & 0.471 & 0.552 & \textbf{0.661} & 0.516 \\
\midrule
\multirow{2}{2.8cm}{$\mathcal{L}_{\text{TOTAL}}$} 
& PSDS1↑ & \textbf{0.154} & \textbf{0.216} & 0.270 & 0.332 & \textbf{0.243} \\
& PSDS2↑ & \textbf{0.397} & \textbf{0.487} & \textbf{0.567} & 0.634 & \textbf{0.521} \\
\bottomrule
\end{tabular}
\begin{tablenotes}
\scriptsize
\item[a] $\mathcal{L}_{\text{BASELINE}} = \mathcal{L}_{\text{SED}} + \mathcal{L}_{\text{SED-CON}}$
\item[b] AVG. represents the average results from different SNR levels.
\end{tablenotes}
\end{threeparttable}
\vspace{-18pt}
\end{table}

\begin{table*}[t]
\centering
\resizebox{0.8\textwidth}{!}{
\begin{threeparttable}
\caption{Performance comparison of different systems under various conditions}
\label{tab:comparison}
\renewcommand{\arraystretch}{1.3}
\setlength{\tabcolsep}{8pt}
\small
\begin{tabular}{@{}lccccccc@{}}
\toprule
ID & System & Clean & WD@-5dB & WD@0dB & WD@+5dB & WD@+10dB & AVG. \\ 
\midrule
M1 & CRNN \cite{11011161} & 1.274 & 0.255 (80.0\%↓) & 0.445 (65.1\%↓) & 0.691 (45.8\%↓) & 0.920 (27.8\%↓) & 0.717 \\ 
M2 & SOD-SED \cite{guan2024sound} $\spadesuit$ & 1.254 & 0.226 (82.0\%↓) & 0.410 (67.3\%↓) & 0.605 (51.8\%↓) & 0.778 (37.9\%↓) & 0.611 \\ 
M3 & FDY-SED \cite{nam2022frequency} $\spadesuit$ & 1.122 & 0.147 (86.9\%↓) & 0.339 (69.8\%↓) & 0.568 (49.4\%↓) & 0.783 (30.2\%↓) & 0.592 \\ 
M4 & ATST-SED \cite{shao2024fine} $\spadesuit$,$\diamondsuit$ & \textbf{1.385} & 0.323 (76.7\%↓) & 0.583 (57.9\%↓) & \textbf{0.839} (39.4\%↓) & \textbf{1.098} (20.7\%↓) & 0.846 \\ 
M5 & LLM-based \cite{11011161} & 1.136 & 0.371 (67.3\%↓) & 0.547 (51.8\%↓) & 0.728 (35.9\%↓) & 0.891 (21.6\%↓) & 0.735 \\ 
M6 & Ours & 1.269 & \textbf{0.551} (56.6\%↓) & \textbf{0.703} (44.6\%↓) & 0.837 (34.0\%↓) & 0.966 (23.9\%↓) & \textbf{0.865} \\ 
\bottomrule
\end{tabular}
\begin{tablenotes}
\small
\item Clean means DESED dataset, and WD@XdB means WildDESED dataset corresponding to X SNR level, ↓ represents decline rate compared with clean data for same model, $\spadesuit$ represents the original authors' model that we reproduced ourselves. $\diamondsuit$ indicates that the pretrained ATST-FRAME encoder has been further fine-tuned using the additional Audioset-2M data.
\end{tablenotes}
\end{threeparttable}
}
\vspace{-18pt}
\end{table*}

\section{Experiments}

\subsection{Datasets}

We evaluate our noise-robust SED system using the DESED \cite{desed} and WildDESED \cite{xiao2024wilddesed} datasets. DESED contains 10-second clips (14.8 hours total) labeled with 10 event classes. WildDESED extends DESED by adding four SNR levels: -5dB, 0dB, +5dB, and +10dB, while keeping other conditions unchanged. All audio is resampled to 16 kHz. For training, we use: (1) 10,000 synthetic strong-labeled clips, (2) 1,578 weak-labeled real clips, and (3) 14,412 unlabeled real clips. To enhance robustness, we add 40,000 synthetic clips from WildDESED with varying SNRs. For evaluation, we follow \cite{11011161}, using the DESED validation set (1,168 real recordings) as clean test data, and the WildDESED validation set to test noise performance across all SNR levels.

\subsection{Baseline System}

We build our system upon the following backbone architecture: Our framework uses a CRNN model \cite{11011161} as the detection network, trained on noisy audio data with BEATs embeddings \cite{chen2023beats} as input features. During testing, we integrate AudioSep-DP \cite{yin2025exploring} as the LASS module. This module first isolates target sound events from noisy mixtures, after which the CRNN performs event detection on the separated signals. This design serves as the common backbone for both our proposed method and the compared systems.

\subsection{Implementation Details}

We adopt the polyphonic sound event detection scores (PSDS) \cite{bilen2020framework} as the evaluation metric. The PSDS provides two different scenarios: scenario 1 (\textit{PSDS1}) emphasizes temporal localization accuracy, whereas scenario 2 (\textit{PSDS2}) focuses on event identification accuracy.

As discussed earlier, unlike \cite{11011161}, our method uses a single unified model for both training and testing. The model is trained for 200 epochs with early stopping, using the Adam optimizer \cite{kingma2014adam} (initial learning rate: 0.001, ramp-up schedule). We apply the Mean Teacher strategy \cite{tarvainen2017mean} with EMA parameter $\alpha=0.999$. Each training batch includes 24 strong, 24 weak, and 48 unlabeled clips (batch size = 96). A median filter with a window size of 7 is used in post-processing. The loss weights are set to \(\rho_{\text{SUP}}=0.2\) and \(\rho_{\text{CON}}=0.012\). We also adopt curriculum learning \cite{bengio2009curriculum} to gradually train on data with increasing SNR levels. During testing, a 0.5 threshold is applied to convert frame-level outputs into clip-level queries for the LASS model. Only the SED branch of the student model is used for evaluation across all SNR levels. 

\begin{figure}[t!]
\centering
\includegraphics[width=0.9\columnwidth]{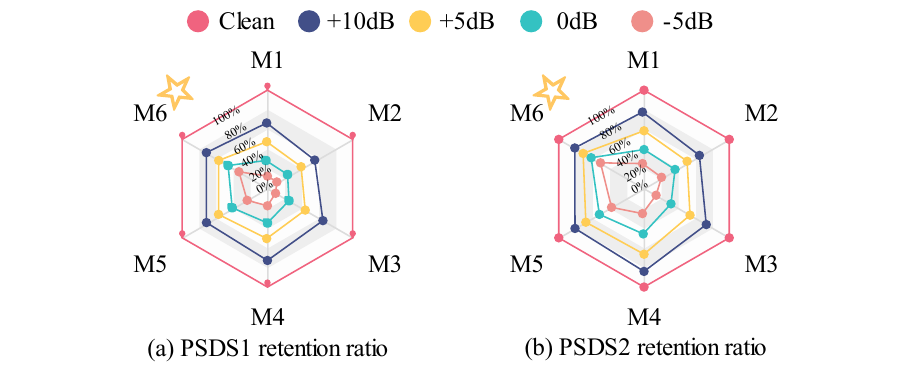}
\vspace{-12pt}
\caption{Illustrates the retention ratio to the systems performance test in clean condition, the closer to the center of radar, the lower the SNR, M1 to M6 represent the ID in Table \ref{tab:comparison}, $\bigstar$ signifies that our proposed framework.}
\vspace{-6mm}
\label{fig:fig3}
\end{figure}

\section{Results and Analyses}

\subsection{Ablation Study}

We conduct ablation studies along two axes: (1) the impact of EAD sub-branches (local and global) and (2) the effect of inter-task consistency constraints. Results are shown in Table~\ref{tab:ablation}.

(1) \textit{EAD sub-branch analysis:} We first evaluate local-EAD by adding \(\mathcal{L}_{\text{local}}\) to the baseline. As shown in the third-to-last row, both PSDS1 and PSDS2 improve, especially at lower SNR levels, showing that local-EAD provides robust guidance and generalizes well to unseen noise. When both global-EAD and local-EAD are used together (\(\mathcal{L}_{\text{EAD}}\)), the performance improves at high SNR, but drops at low SNR, particularly in PSDS2. This suggests that the added complexity from combining sub-branches can lead to redundant signals, weakening EAD’s contribution when the signal is more degraded.

(2) \textit{Inter-task consistency constraints:} We then evaluate the full model with cooperative training ($\mathcal{L}_{\text{TOTAL}}$), shown in the last row. Results show consistent gains in both metrics across all SNR levels. This confirms that inter-task consistency enhances robustness. However, performance slightly declines at high SNR, likely due to train-test mismatch: consistency is enforced during training, but not during testing, where only SED is active. In clean conditions, this coupling can hinder SED, which performs well even without auxiliary tasks. In contrast, at low SNR, the shared features learned through consistency remain helpful, supporting stronger generalization.

\begin{figure}[t!]
\centering
\includegraphics[width=1.0\columnwidth]{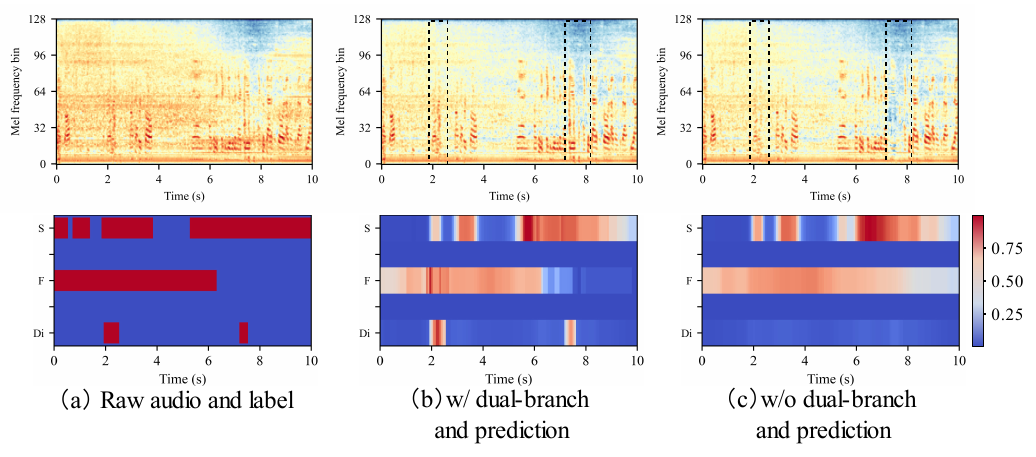}
\vspace{-20pt}
\caption{Separation and prediction visualizations of test example resulted by the proposed dual-branch framework (b) and baseline (c) with corresponding raw audio and label (a).}
\vspace{-4mm}
\label{fig:fig4}
\end{figure}

\subsection{Comparing with Other systems}

We compare our framework with several state-of-the-art methods to assess its effectiveness in noisy environments. Table~\ref{tab:comparison} presents the total PSDS scores (PSDS1 + PSDS2) across varying SNR levels. Methods M1–M4 show strong performance in clean conditions but degrade in noisy environments, as they focus mainly on representation or timestamps alignment. Although ATST-SED excels at high SNR due to additional pre-training on AudioSet-2M \cite{gemmeke2017audio}, its generalization drops quickly at lower SNRs. For instance, at -5dB, our system outperforms SOD-SED by more than doubles in total PSDS, indicating stronger robustness. Compared to the LLM-based system (M5), which uses a two-stage training pipeline and double the model parameters, our method is end-to-end, more efficient, and performs better on average across all SNRs. As shown in Fig.~\ref{fig:fig3}(a–b), our system retains performance more effectively than all baselines as SNR decreases, confirming the superior noise robustness of our cooperative dual-branch framework with LASS.

\subsection{Visualization}

To further validate the effectiveness of our collaborative dual-branch framework in noisy SED and its impact on LASS performance, we analyzed a test clip at 0dB SNR, which includes three target sounds: Speech (S), Frying (F), and Dishes (Di). As shown in Fig.\ref{fig:fig4}(b–c), our method enables the LASS model to separate sounds more accurately, due to improved text-query quality driven by global-EAD's event count estimation. Compared to the baseline, our system better matches the ground truth (Fig.\ref{fig:fig4}(a)) in both event classification and timestamps alignment, confirming the value of combining global and local EAD with SED training. These findings demonstrate the robustness and precision of our approach in challenging noisy conditions.

\section{Conclusion}

The cooperative dual-branch framework, specifically proposed for SED under noisy conditions, is presented in this letter. This framework leverages consistency between EAD and SED to enhance temporal localization at the frame-level and generate reliable text-queries for LASS model at the clip-level. Experimental results demonstrate that our method improves performance across various SNR levels, confirming its effectiveness. The proposed approach also shows remarkable potential for addressing noise-related challenges in diverse audio processing applications.

\bibliographystyle{IEEEtran}
\clearpage
\balance
\bibliography{reference}
\end{document}